\documentclass[amsmath,amssymb,numbers]{revtex4}

\begin{document}

\title{Geometrical Underpinning of Finite Dimensional Hilbert space}

\author{M. Revzen}
\affiliation {Department of Physics, Technion - Israel Institute of Technology, Haifa
32000, Israel}

\date{\today}

\begin{abstract}

Finite geometry is employed to underpin operators in finite, d, dimensional Hilbert space.
The central role of mutual unbiased bases (MUB) states projectors is exhibited.
Interrelation among operators in Hilbert space, revealed through their (finite) dual affine
plane geometry (DAPG) underpinning is studied. Transcription to (finite) affine plane
geometry (APG) is given and utilized for their interpretation.
\end{abstract}


\maketitle

\section{Introduction}

Several recent studies \cite{wootters4,saniga,planat1,planat2,combescure} consider the
relation between finite geometry and finite dimensional Hilbert space. Such relation, aside
from its intrinsic interest, reveals interrelation among physical operators that may be
hidden otherwise. The present work emphasizes a particular branch of finite geometry: dual
affine plane geometry (DAPG) \cite{wootters4,shirakova,tomer,diniz}. The physical operators
that play the central role in this work are the mutual unbiased bases (MUB) state's
projectors \cite{wootters1,wootters2,bengtsson,amir}. The association of MUB projectors
with affine plane geometry (APG) was treated in depth in \cite{wootters4} who emphasized
the relevance of the trace in these studies. In the present work we are concerned with
underpinning of the physical operators with both geometries and indicate the inter
transitions between DAPG and APG. Our approach allows the definition of mappings of Hilbert
space operators onto the lines and points of the geometries in addition to exhibiting the
geometry based interrelation among operators.\\
The finite dimensional characteristics of both affine plane geometry (APG) and its dual
(DAPG) are listed in section II. We describe therein the interrelations among points and
lines required for a realization of the geometries. Mutual unbiased bases (MUB) whose
state's projectors are central to our study are defined and discussed briefly in the next
section, section III. In section IV we present our underpinning scheme of DAPG for Hilbert
space operators. The section contains some geometrically based interrelations among
operators. In section V we study further the so called line operators i.e. operators
underpinned with DAPG lines. We give here the equation governing the DAPG lines and derive
some explicit properties of of the operators. APG is considered in section VI where we
discuss the DAPG and APG underpinned operators. In section VII we briefly indicate the
geometrically assisted mappings of operators in Hilbert space onto lines and points of the
geometry. The last section, section VIII, contained brief summary and discussions.

 \section{   Finite Geometry and Hilbert Space Operators}

We now briefly review the essential features of finite geometry required for our study
\cite{grassl,diniz,shirakova,tomer,wootters4}.\\
A finite plane geometry is a system possessing a finite number of points and lines. There
are two kinds of finite plane geometry: affine and projective. We shall confine ourselves
to affine plane geometry (APG) which is defined as follows. An APG is a non empty set
whose
elements are called points. These are grouped in subsets called lines subject to:\\
1. Given any two distinct points there is exactly one line containing both.\\
2. Given a line L and a point S not in L ($S \ni L$), there exists exactly one line L'
containing S
such that $L \bigcap L'=\varnothing$. This is the parallel postulate.\\
3. There are 3 points that are not collinear.\\
It can be shown \cite{diniz,shirakova} that for $d=p^m$ (a power of prime) APG can be
constructed (our study here is for d=p) and the following properties are, necessarily,
built in: \\
a. The number of points is $d^2;$ $S_{\alpha},\;{\alpha = 1,2,...d^2}$ and the number of
lines is d(d+1); $L_j,\;j=1,2....d(d+1)$.\\
b. A pair of lines may have at most one point in common: $L_j\bigcap
L_k=\lambda;\;\lambda=0,1\;for j \ne k$.\\
c. Each line is made of d points and each point is common to d+1 lines:
$L_j=\bigcup_{\alpha}^d S_{\alpha}^j$, $S_{\alpha}=\bigcap_{j=1}^{d+1}L_j^{\alpha}.$\\
d. If a line $L_j$ is parallel to the distinct lines $L_k\;and\;L_i$ then $L_k \parallel
L_i$. The $d^2$ points are grouped in sets of d parallel lines. There are d+1 such
groupings.\\
e. Each line in a set of parallel lines intersect each line of any other set:
$L_j\bigcap L_k=1;\;L_j \nparallel L_k.$ \\
The above items will be referred to by APG (x), with x=a,b,c,d or e.\\

The existence of APG implies \cite{diniz,grassl,shirakova}the existence of its dual
geometry DAPG wherein the points and lines are interchanged. Since we shall study
extensively this, DAPG, we list the
corresponding properties for it. We shall refer to these by DAPG(y):\\
a. The number of lines is $d^2$, $L_j,\;j=1,2....d^2.$ The number of points is d(d+1),
$S_{\alpha},\;{\alpha = 1,2,...d(d+1)}.$\\
b. A pair of points on a line determine a line uniquely. Two (distinct) lines share one and only
one point.\\
c. Each point is common to d lines. Each line contain d+1 points.\\
d. The d(d+1) points may be grouped in sets, $R_{\alpha}$, of d points each no two of a set
share a line. Such a set is designated by $\alpha' \in \{\alpha \cup M_{\alpha}\},\;
\alpha'=1,2,...d$. ($M_{\alpha}$ contain all the points not connected to $\alpha$ - they
are not connected among themselves.) i.e. such a set contain d disjoined (among themselves)
points. There are d+1 such sets:
\begin{eqnarray}
\bigcup_{\alpha=1}^{d(d+1)}S_{\alpha}&=&\bigcup_{\alpha=1}^d R_{\alpha};\label{a}\\
R_{\alpha}&=&\bigcup_{\alpha'\epsilon\alpha\cup M_{\alpha}}S_{\alpha'};\label{b}  \\
R_{\alpha}\bigcap R_{\alpha'}&=&\varnothing,\;\alpha\ne\alpha'.\label{c}
\end{eqnarray}
e. Each point of a set of disjoint points is connected to every other point not in its
set.\\

\section{   Finite dimensional Mutual Unbiased Bases, MUB, Brief Review}

In a finite, d-dimensional, Hilbert space two complete, orthonormal vectorial bases, ${\cal
B}_1,\;{\cal B}_2$,
 are said to be MUB if and only if (${\cal B}_1\ne {\cal B}_2)$
 \cite{wootters1,wootters2,wootters3,wootters4,tal,vourdas,klimov2,bengtsson,planat1,amir,mello1,ent,peres}

\begin{equation}
\forall |u\rangle,\;|v \rangle\; \epsilon \;{\cal B}_1,\;{\cal B}_2 \;resp.,\;\;|\langle
u|v\rangle|=1/\sqrt{d}.
\end{equation}
The physical meaning of this is that knowledge that a system is in a particular state in
one basis implies complete ignorance of its state in the other basis.\\
Ivanovic \cite{ivanovich} proved that there are at most d+1 MUB, pairwise, in a
d-dimensional Hilbert space and gave an explicit formulae for the d+1 bases in the case of
d=p (prime number). Wootters and Fields \cite{wootters2} constructed such d+1 bases for
$d=p^m$ with m an integer. Variety of methods for construction of the d+1 bases for $d=p^m$
are now available
\cite{tal,klimov2,vourdas}. Our present study is confined to $d=p\;\ne2$.\\
 We now give explicitly the MUB states in conjunction with the algebraically complete
 operators \cite{schwinger,amir} set:
 $\hat{Z},\hat{X}$.  Thus we label the d distinct states spanning the Hilbert space,
 termed
 the computational basis, by $|n\rangle,\;\;n=0,1,..d-1; |n+d\rangle=|n\rangle$
\begin{equation}
\hat{Z}|n\rangle=\omega^{n}|n\rangle;\;\hat{X}|n\rangle=|n+1\rangle,\;\omega=e^{i2\pi/d}.
\end{equation}
The d states in each of the d+1 MUB bases \cite{tal,amir}are the states of computational
basis (CB) and
\begin{equation} \label{mxel}
|m;b\rangle=\frac{1}{\sqrt
d}\sum_0^{d-1}\omega^{\frac{b}{2}n(n-1)-nm}|n\rangle;\;\;b,m=0,1,..d-1.
\end{equation}
Here the d sets labeled by b are the bases and the m labels the states within a basis. We
have \cite{tal}
\begin{equation}\label{tal1}
\hat{X}\hat{Z}^b|m;b\rangle=\omega^m|m;b\rangle.
\end{equation}
For later reference we shall refer to the computational basis (CB) by b=-1. Thus the above
gives d+1 bases, b=-1,0,1,...d-1 with the total number of states d(d+1) grouped in d+1 sets
each of d states. We have of course,
\begin{equation}\label{mub}
\langle m;b|m';b\rangle=\delta_{m,m'};\;\;|\langle m;b|m';b'\rangle|=\frac{1}{\sqrt d},
\;\;b\ne b'.
\end{equation}
We remark at this junction that the eigen values of the CB might be considered finite
dimensional modulated position values ("q") and the eigenvalues of shifting operator, X,
modulated momentum ("p").\\
This completes our discussion of MUB.\\

\section{ DAPG underpinning of d-dimensional Hilbert space}

We first list some direct consequences of DAPG.
DAPG(c) allows the definition;
\begin{equation}\label{A}
S_{\alpha}=\frac{1}{d}\sum_{j\in\alpha}^{d} L_j.
\end{equation}
This implies,
\begin{equation}
\sum_{\alpha' \in \alpha \cup M_{\alpha}}^{d}S_{\alpha'}=\frac{1}{d}\sum^{d^2} L_j,
\end{equation}
leading via DAPG(d) to
\begin{equation}
\sum_{\alpha}^{d+1} \sum_{\alpha' \in \alpha \cup M_{\alpha}}^{d}S_{\alpha'}=\sum^{d(d+1)}S_{\alpha}
                          =\frac{d+1}{d}\sum^{d^2} L_j.
\end{equation}

We thus have,
\begin{equation}\label{I}
\sum_{\alpha' \in \alpha \cup M_{\alpha}}^{d}S_{\alpha'}=\frac{1}{d}\sum^{d^2} L_j=
\frac{1}{d+1}\sum^{d(d+1)}S_{\alpha}.
\end{equation}

 Now the underpinning of Hilbert space operators with DAPG  will be undertaken.
We consider d=p, a prime. For d=p we may construct d+1 MUB
\cite{ivanovich,tal,wootters1,wootters4}. Points will be associated with MUB state projectors.
To this end we recall that we designate the MUB
states by $|m,b\rangle.$ with $b=0,1,2...d-1$ labels the eigenfunction of, resp. $XZ^b$. m labels the state within a basis. We designate the computational basis, CB, by  b=-1. The projection operator defined by,
\begin{equation}\label{ptop}
\hat{A}_{\alpha}\equiv |m,b\rangle \langle
b,m|;\;\alpha=\{b,m\};\;\;b=-1,0,1,2...d-1;\;m=0,1,2,..d-1.
\end{equation}
The point label, $\alpha=(m,b)$ is now associated with the projection operator,
$A_{\alpha}$ . We now consider a realization, possible for d=p, a prime , of a d
dimensional DAPG,  as points marked on a rectangular whose horizontal width (x-axis) is made of d+1
columns of points. Each column is labeled by b, and its vertical height (y axis) is made of d points
each marked with m. The total number of points is d(d+1) - there are d points in each of
the d+1 columns. We associate the d points $m=0,1,2,...d-1$ in each set, labeled by b,
($\alpha \sim (m,b)$) to the {\it disjointed} points of DAPG(d), $R_{\alpha}$ viz. for fixed
b $\alpha' \in \alpha \cup M_{\alpha}$ form a column. The columns are arranged according to
their basis label, b. The first being b=-1, $\alpha_{-1}=(m,-1);m=0,1,...d-1,$ pertains to
the computational basis (CB). Lines are now made of d+1 points, each of different b.
A point $S_{\alpha}$  underpins a Hilbert space state projector, $A_{\alpha}$.
i.e. $A_{\alpha}^{2}=A_{\alpha},$ and $trA_{\alpha}=1.$
We designate the line operator underpinned with $L_j,$ by $P_j$. Thus the above relations now hold with
$S_{\alpha}\leftrightarrow A_{\alpha};\;\;L_j\leftrightarrow P_j$.\\

Now DAPG(c) (and Eq.(\ref{ptop}),({\ref{I}))implies that
$A_{\alpha};\;\alpha=0,1,2...d-1;\;\;\alpha \in \alpha' \cup M_{\alpha'}$ forms an
orthonormal basis for the d-dimensional Hilbert space:
\begin{eqnarray}
\sum_m^d |m,b\rangle\langle b,m|&=&\sum_{\alpha' \in \alpha\cup M_{\alpha}}^d\hat{A}_{\alpha'}
=\hat{I} \nonumber \\
\sum_{\alpha}^{d(d+1)}\hat{A}_{\alpha}&=&(d+1)\hat{I}.
\end{eqnarray}

e.g. for d=3 the underpinning's schematics is,
\[ \left( \begin{array}{ccccc}
m\backslash b&-1&0&1&2 \\
0&A_{(0,-1)}&A_{(0,0)}&A_{(0,1)}&A_{(0,2)}\\
1&A_{(1,-1)}&A_{(1,0)}&A_{(1,1)}&A_{(1,2)}\\
2&A_{(2,-1)}&A_{(2,0)}&A_{(2,1)}&A_{(2,2)}\end{array} \right)\].\\

Eq.(\ref{A}) implies,
\begin{equation}\label{A1}
A_{\alpha}=\frac{1}{d}\sum_{j\in\alpha}^d P_j.
\end{equation}

Evaluating
\begin{equation}
\sum_{\alpha\in j}A_{\alpha}=\frac{1}{d}\sum_{\alpha\in j}\sum_{j'\in\alpha}P_j=
\frac{1}{d}\big[\sum^{(d-1)(d+1)}_{j'\ne j} P_{j'}+(d+1)P_j\big]=I+P_j.
\end{equation}
i.e.,
\begin{equation}\label{P}
P_j=\sum^{d+1}_{\alpha\in j}A_{\alpha}-I.
\end{equation}

 Eq.(\ref{mub}) implies,
 \begin{equation}
 tr A_{\alpha}A_{\alpha'}=\begin{cases}1;\;\;\alpha=\alpha'\\
 0;\;\;\alpha \ne \alpha';\alpha \in \alpha'\cup M_{\alpha'}.\\
 \frac{1}{d};\;\alpha\ne \alpha';\;\alpha \ni \alpha'\cup M_{\alpha'}.\end{cases}
 \end{equation}

Hence, using Eq.(\ref{A}),(\ref{mub}),
\begin{equation}\label{delta1}
tr A_{\alpha}P_j=\begin{cases}\sum_{\alpha' \ne \alpha}^{d}tr A_{\alpha}A_{\alpha'}=1;\;\alpha\in j.\\ \sum_{\alpha'\ne \alpha}^{d}tr A_{\alpha}A_{\alpha'}-A_{\alpha}=0;\;\alpha\ni j. \end{cases}
\end{equation}
Trivially
\begin{equation}
trP_j=\sum^{d+1}trA_{\alpha}-1=1.
\end{equation}
\begin{equation}
trP_jP_{j'}=\sum_{\alpha'\in j'}trP_jA_{\alpha'}-1=\begin{cases}d\;\;j=j'\\0\;\;j\ne j',\end{cases}
\end{equation}
i.e.
\begin{equation}
trP_jP_{j'}=d\delta_{j,j'}.
\end{equation}

An alternative view of the Lambda function is gained via

\begin{equation}\label{delta2}
tr A_{\alpha}P_j=\frac{1}{d}\sum^{d}P_{j'}P_j=\begin{cases}\frac{1}{d}\big(tr P_j^2+tr \sum_{j'\ni\alpha} P_{j'}P_j\big)=1;\;\;j\in \alpha \\
\sum_{j'\ne j}P_{j'}P_j=0;\;\;j\ni \alpha.\end{cases}
\end{equation}
Note that the case of $j\ni\alpha$ {\it implies} $j\in M_{\alpha}$.\\
These are summarized by

\begin{equation}\label{del1}
tr A_{\alpha}P_j=\begin{cases}1;\;\alpha\in j,\\
0;\;\alpha\ni j, \end{cases}
\end{equation}
and
\begin{equation}\label{del2}
tr A_{\alpha}P_j=\begin{cases}1;\;\;j\in \alpha \\
0;\;\;j\ni \alpha.\end{cases}
\end{equation}

\section{  Geometric Underpinning of MUB Quantum Operators: The line operator}

We now consider a particular realization of DAPG of dimensionality $d=p,\ne 2$ which is the
basis of our present study. We arrange the aggregate the d(d+1) points, $\alpha$, in a
$d\cdot(d+1)$matrix like rectangular array of d rows and d+1 columns. Each column is made
of a set of d points  $R_{\alpha}=\bigcup_{\alpha'\epsilon\alpha\cup
M_{\alpha}}S_{\alpha'};$  DAPG(d). We label the columns by b=-1,0,1,2,....,d-1 and the rows
by m=0,1,2...d-1.( Note that the first column label of -1 is for convenience and does not
designate negative value of a number.)  Thus $\alpha=m(b)$ designate a point by its row, m,
and its column, b; when b is allowed to vary - it designate the point's row position in
every column. We label the left most column by b=-1 and with increasing values of b, the
basis label, as we move to the right. Thus the right most column is b=d-1. We now assert
that the d+1 points, $m_j(b), b=0,1,2,...d-1,$  and  $m_j(-1)$, that form the line j which
contain the two (specific) points m(-1) and m(0) is given by (we forfeit the subscript j -
it is implicit),
\begin{eqnarray}\label{m(b)}
m(b)&=&\frac{b}{2}(c-1)+m(0),\;mod[d]\;\;b\ne -1, \nonumber \\
m(-1)&=&c/2.
\end{eqnarray}

The rationale for this particular form will be clarified below. Thus a line j is
parameterized fully by $j=(m(-1),m(0))$. (Note: since b takes on the values -1 and 0 in our line labeling a more economic label for j is $j=(m_{-1},m_0)$ i.e. the m values for b=-1 and 0. We shall use either when no confusion should arise.) We now prove that the set $j=1,2,3...d^2$ lines
covered by Eq.(\ref{m(b)}) with the points as defined above form a DAPG.\\
\noindent 1. Since each of the  parameters, m(-1) and m(0), can have d values the number of
lines is $d^2$; the number of points in a line is evidently d+1: a point for each b.  DAPG(a).\\
\noindent 2. The linearity of the equation precludes having two points with a common value
of b on the same line, DAPG(d). Now consider two points on a given line, $m(b_1),m(b_2);\;b_1\ne
b_2$. We have from Eq.(\ref{m(b)}), ($b\ne -1,\;b_1 \ne b_2$)
\begin{eqnarray}\label{twopoints}
m(b_1)&=&\frac{b_1}{2}(c-1)+m(0),\;\;mod[d]\nonumber\\
m(b_2)&=&\frac{b_2}{2}(c-1)+m(0),\;\;mod[d].
\end{eqnarray}
These two equation determine uniquely ({\it for d=p, prime}) m(-1) and m(0). DAPG(b).\\
\noindent For fixed point, m(b), $c\Leftrightarrow m(0)$ i.e the number of free parameters
is d (the number of points on a fixed column). Thus each point is common to d lines. That
the line contain d+1 is obvious. DAPG(c).\\
\noindent 3. As is argued in 2 above no line contain two points in the same column (i.e.
with equal b). Thus the d points, $\alpha,$ in a column form a set
$R_{\alpha}=\bigcup_{\alpha'\epsilon\alpha\cup M_{\alpha}}S_{\alpha'},$ with trivially
$R_{\alpha}\bigcap R_{\alpha'}=\varnothing,\;\alpha\ne\alpha',$ and
$\bigcup_{\alpha=1}^{d(d+1)}S_{\alpha}=\bigcup_{\alpha=1}^d R_{\alpha}.$ DAPG(d).\\
\noindent 4. Consider two arbitrary points {\it not} in the same set, $R_{\alpha}$ defined
above: $m(b_1),\;m(b_2)\;\;(b_1\ne b_2).$ The argument of 2 above states that, {\it for
d=p}, there is a unique solution for the two parameters that specify the line containing
these points. DAPG(e).\\
We illustrate the above for d=3, where we explicitly specify the points contained in the
line $j=\big(m(-1)=(1,-1),m(0)=(2,0)\big)$
\[ \left( \begin{array}{ccccc}
m\backslash b&-1&0&1&2 \\
0&\cdot&\cdot&\cdot&(0,2)\\
1&(1,-1)&\cdot&(1,1)&\cdot\\
2&\cdot&(2,0)&\cdot&\cdot\end{array} \right)\].\\
For example the point m(1) is gotten from
$$ m(1)= \frac{1}{2}(2-1)+2=1\;\;mod[3]\;\;\rightarrow\;m(1)=(1,1).$$
Similar calculation gives the other point: m(2)=(0,2). i.e. the line j=(1,2) contains the points
(1,-1),(2,0),(1,1) and (0,2).

The geometrical line, $L_j, \;j=(1,2)$ given above upon being transcribed
to its operator formula is via Eq.(\ref{P}),

\begin{equation}\label{lineop}
P_{j=(1,2)}=A_{(1,-1)}+A_{(2,0)}+A_{(1,1)}+A_{(0,2)}-\hat{I}.
\end{equation}

Evaluating the point operators, $\hat{A}_{\alpha}$,
\begin{equation}\label{point1}
A_{(1,-1)}=\begin{pmatrix}0&0&0\\0&1&0\\0&0&0\end{pmatrix},A_{(2,0)}=\frac{1}{3}\begin{pmatrix}1&\omega^2&\omega\\
\omega&1&\omega^2\\\omega^2&\omega&1\end{pmatrix},A_{(1,1)}=\frac{1}{3}\begin{pmatrix}1&\omega&\omega\\\omega^2&1&1\\
\omega^2&1&1\end{pmatrix},A_{(0,2)}=\frac{1}{3}\begin{pmatrix}1&1&\omega\\1&1&\omega\\\omega^2&\omega^2&1\end{pmatrix},
\end{equation}
and evaluating the sum, Eq.(\ref{lineop}), gives
\begin{equation} \label{pj}
P_{j:(m(-1)=1,m(0)=2)}=\begin{pmatrix}0&0&\omega\\ 0&1&0\\ \omega^2&0&0\end{pmatrix}.
\end{equation}

This operator obeys $P_{(1,2)}^2=\hat{I}$. We shall now show that this is quite general, viz $P_j^2=\hat{I},\;\forall j$.

Returning to Eqs.(\ref{ptop},\ref{mxel}), these equations imply that, the projection operators $A_{\alpha}$, in the CB representation are given by,

\begin{equation}\label{point}
\big(A_{\alpha=m,b}\big)_{n,n'}=\begin{cases}
\frac{\omega^s}{d};\;\;s=(n-n')(\frac{b}{2}[n+n'-1]-m),\;\;
b\ne-1,\\
\delta_{n,n'}\delta_{c/2,n}\;\;b=-1.\end{cases}
\end{equation}

Consider two distinct columns, b,b' $(b,b'\ne-1)$ and given the matrix elements n,n' $(n\ne
n')$ of a projector $\big(A_{\alpha=m,b}\big)_{n,n'},$ compare it with
$\big(A_{\alpha'=m',b'}\big)_{n,n'}.$ If s (Eq.(\ref{point})) is $\ne s'$ i.e.
$\frac{b}{2}(n+n')-1)-m \ne \frac{b'}{2}(n+n')-1)-m'$ pick another projector in the same
column, b' (i.e vary m'). Since m' = 0,1,2...d-1 there is one (and only one)
$\big(A_{\alpha'=m',b'}\big)_{n,n'}$ such that
$\big(A_{\alpha=m,b}\big)_{n,n'}=\big(A_{\alpha'=m',b'}\big)_{n,n'}.$ Now consider another
matrix element $\big(A_{\alpha}\big)_{\bar{n},\bar{n}'}.$ We have trivially that
$\big(A_{\alpha}\big)_{\bar{n},\bar{n}'}=\big(A_{\alpha'}\big)_{\bar{n},\bar{n}'}$ iff
$\bar{n}+\bar{n}'=n+n'$. i.e. all matrix elements (n,n') with n+n'=c (constant) are such
that $\big(A_{\alpha}\big)_{n,n'}=\big(A_{\alpha'}\big)_{n,n'}.$ These elements are
situated along a line perpendicular to the diagonal of the matrices. We refer to this
perpendicular as FV (foliated vector), it is parameterized by  c.\\
We now assert that all other (non diagonal) matrix elements are unequal. i.e. for $b \ne
b'$, $\big(A_{\alpha=m,b} \big)_{n,n'}$ $\ne\;\big(A_{\alpha'=m',b'} \big)_{n,n'},\;\forall
\;n,n' \ni FV.$ Proof: Let two elements n,n' and l,l' with $n\ne n';\;l\ne l'$ in the two
matrices be equal. Thus (c=n+n', c'=l+l'):
\begin{eqnarray}
\frac{b}{2}(c-1)-m&=&\frac{b'}{2}(c-1)-m', \;\;\;and\nonumber \\
\frac{b}{2}(c'-1)-m&=&\frac{b'}{2}(c'-1)-m',\;\; \nonumber \\
\end{eqnarray}
These {\it imply} c=c', QED.  Now consider s=0. Then all the matrix elements along FV are
1/d. We have then that for $\big(A_{\alpha} \big)_{n,n'}=\big(A_{\alpha'}
\big)_{n,n'}=\omega^s/d,\;s \ne 0,$ d-1
matrix elements along FV are all distinct. The diagonal is common to all.\\
We have thus a prescription for d projectors, $A_{(m,b)},$ one for each b, ($b \ne -1),$
all having equal matrix elements along FV labelled by c. We supplement these with the
projector $A_{(c/2,-1)}=|c/2\rangle\langle c/2|$ to have the d+1 "points" constituting a
line j . $(|c/2 \rangle$ being a state in the CB.) Thus our line is formed as follows: It
emerges from $A_{(c/2,-1)}$ continues to $A_{(m(0),0)}$ in the b=0 column. Then it
continues to the points $A_{m(b),b}$ in succession: b=1,2...d-1 with m(b) determined by
$$\frac{b}{2}(c-1)-m(b)=\frac{b+1}{2}(c-1)-m(b+1).$$ Thus the two parameters, c=2m(-1) and m(1),
determine the line i.e. j=(m(-1),m(1)). The general formula for the line, Eq.(\ref{m(b)},
 now acquires a meaning in terms of the point operators, $A_{\alpha=m(b),b}$.
 The discussion of the properties of the line thus defined confirm that these lines form a realization
 of  DAPG
 lines. The analysis above indicate that the line operator, $P_j$, may be labelled by
 two indices
 $P_{j=(m(-1),m(0))}.$ We now list some
 important consequences of this. We have shown that the matrix elements along a FV direction are
the same for all the point operators $A_{\alpha \in j}.$ Indeed that is how we defined our
lines. On the other hand we argued that the matrix elements {\it not} along the FV are all
distinct. Thence summing up d such terms residing on a fixed line $P_j$ ({\it excluding the
b=-1 and the diagonal term}) sums up for each matrix  element n,n' the d distinct roots of
unity for matrix elements {\it not on FV}, hence for all c,
\begin{equation}
\big(\sum_{\alpha \in j, \alpha \ni
\alpha_{-1}}^{d}\hat{A}_{\alpha}-\hat{I}\big)_{n,n'}=0;\;n,n'\ni
n+n'=c;\;\;\alpha_{-1}=|c/2\rangle\langle c/2|.
\end{equation}
Thus
 $\big(\hat{P}_j\big)_{n,n'}=\big(\sum_{\alpha\in j}^{d+1}\hat{A}_{\alpha}-\hat{I}\big)_{n,n'}\ne 0$
 {\it only} along FV, and is 1 along the diagonal at c/2=m(-1). The sum over $\alpha \in j$ of the matrix
 elements on a FV, which are equal for all  $\hat{A}_{\alpha \in j, \ne -1},$ simply cancel
 the $1/d$. This is illustrated in Eqs.({point1}),({pj}),

Quite generally,
\begin{equation}\label{pc}
(P_{j=m(-1),m(0)})_{n,n'}=\begin{cases}\omega^{-(n-n')m(0)}\delta_{\{(n+n'),2m(-1)\}}\\
0\;\;otherwise. \end{cases}
\end{equation}
Thus,

\begin{equation}\label{p2}
(\hat{P}_{j=m(-1),m(0)}^2)_{n,n'}=\delta_{n,n'}.\;\;i.e.\;\hat{P}_j^2=\hat{I}\;\forall j .
\end{equation}
In appendix A we show that $\hat{P}_j^2=\hat{I}\;\forall j $ implies the operator relation,
$$\sum_{\alpha \ne\alpha' \in
j}\hat{A_\alpha}\hat{A_\alpha'}=\sum_{\alpha \in j}\hat{A_\alpha}.$$

\section{Affine Plane Geometry (APG)}

We now recast our DAPG underpinning into an AFG one. This is achieved by interchanging
lines with points. For notational convenience we refer to m(-1) and m(0) of the DAPG (cf.
Eq.()) by $\xi$ and $\eta$ respectively in their APG image. Thus a line in DAPG
j=(m(-1),m(0)) is a point $(\xi,\eta)$ in its APG image, $\xi,\eta = 0,1,...d-1.$ We now
construct the following realization of a d (=prime) dimensional APG: consider $d\cdot d$
points arranged in a square array of d columns and d rows. Each point is an image of DAPG
line and is specified by $(\xi,\eta)$. e.g. the point whose coordinates are $\xi=1,\eta=2$
represents the line, $L_{(1,2)}$, which in turn underpins the line operator $P_{(1,2)}$
considered above. Now consider "straight" lines in this array defined by
\begin{eqnarray}
\eta&=&r\xi+s \;mod[d]\;\;r,s=0,1,...d-1 \nonumber \\
\xi&=&s'\;mod[d]\;s'=0,1,...d-1.
\end{eqnarray}
We contend that these points and lines form a realization of APG:\\
\noindent a. The number of points is evidently $d^2$ and the number of lines $d^2+d=d(d+1)$.\\
\noindent b. The linearity of the equations guarantee that all non parallel lines share one point and
one has d+1 sets each with d parallel lines.\\
\noindent c. Each line is made of d points and each point is common to d+1 lines, e.g. the point
$\xi_1,\eta_2$ is shared by the lines
$$\eta_2=r\xi_1+s\;and \;\eta_2=r"\xi_1+s"\;with\;(r-r")\xi_1=s-s" \;mod[d]\;s'=\eta_2\;mod[d].$$
The uniqueness of the solutions are assured by the d=prime requirement.\\
\noindent d. s=0,1,...d-1 and s'=0,1,...d-1 divide the lines to sets of d mutually parallel lines each.\\
\noindent e. The linearity of the equations assures us that each line share one point with each line
not belonging to its class of mutually parallel lines.\\
The consistency of our considerations requires  that setting the image of DAPG line as an
APG point implies that the image of APG line be a DAPG point. This consistency is
demonstrated  upon proving the following proposition: The d lines images within DAPG of the
d points residing on a APG line share a
point. The proof is as follows:\\
\noindent 1. The d points $\xi=0,1,...d-1$ on the line $\xi=s'$ trivially share the point
m(-1)=s'. Thus they are the images of the d lines "emerging" from the point s' in the
column b=-1 of the DAPG realization.\\
\noindent 2. Consider two arbitrary APG points $(\eta',\xi')$ and $(\eta",\xi")$ that lie
on the line $\eta=r\xi+s,\;mod[d]$ For their lines DAPG images to share a point we must
have, Eq.(), $b/2(2\xi'-1)+r\xi'+s$ equals $b/2(2\xi"-1)+r\xi"+s,\;mod[d].$ i.e. b+r=0,
mod[d]. Thus their common point is $m(b)=-r/2(2\xi'-1)+r\xi'+s,$ $\rightarrow$
$m(b=-r)=-r/2+s$ i.e. in the column b=-r;mod[d],  row m=s+r/2;mod[d]. This is true for all
the the d points on that line. QED.\\
The consistency requirement is thus fulfilled: Consider APG line (e.g.$ \eta=r\xi+s$). The
APG {\it points} on this line $(\xi',\eta'),(\xi",\eta")...(\xi^d,\eta^d),$ are the images
of DAPG {\it lines}
$[j'=(\xi'=m'(-1),\eta'=m'(0)],[j"=(\xi"=m"(-1),\eta"=m"(0)]...[j^d=..].$\\

Thus the interrelation among {\it the operators}, $A_{\alpha}$ and $P_j$ are identical whether given within DAPG or APG. e.g. Eq.()
$$A_{\alpha}=\frac{1}{d}\sum_{j\in\alpha}^d P_j$$
is, within, DAPG gives the {\it point} operator $A_{\alpha}$ in terms of the d {\it line} operators $P_j$ - while within APG this very same equation gives the very same Hilbert space operator, $A_{\alpha}$, now a {\it line} operator in terms of the d {\it point} operators $P_j$.

As an example let us consider for d=3 the APG line $\eta=\xi+1.$  The APG points on
this line are (0,1),(1,2) and (2,0) reflecting DAPG line operators, cf Eq.(\ref{pc}),
\begin{equation}
P_{(0,1)}=\begin{pmatrix}1&0&0\\ 0&0&\omega\\ 0&\omega^2&0\end{pmatrix};\;\;
   P_{(1,2)}=\begin{pmatrix}0&0&\omega\\ 0&1&0\\ \omega^2&0&0\end{pmatrix};\;\;
P_{(2,0)}=\begin{pmatrix}0&0&\omega\\ 0&1&0\\ \omega^2&0&0\end{pmatrix}.
\end{equation}
Now, Eq.(\ref{trans}) relates these to the DAPG point (i.e. the MUB projector) $|0,2\rangle
\langle2,0|=A_{(0,2)}$
\begin{equation}
\frac{1}{3}\Big[\begin{pmatrix}1&0&0\\ 0&0&\omega\\
0&\omega^2&0\end{pmatrix}+\begin{pmatrix}0&0&\omega\\ 0&1&0\\ \omega^2&0&0\end{pmatrix}+
\begin{pmatrix}0&0&\omega\\ 0&1&0\\
\omega^2&0&0\end{pmatrix}\Big]=\frac{1}{3}\begin{pmatrix}1&\omega^2&\omega\\
\omega&1&\omega^2\\ \omega^2&\omega&1\end{pmatrix},
\end{equation}
where the last matrix is $A_{(0,2)},$ cf. Eq.(\ref{point1}).\\
This completes our discussion of geometrical underpinning of finite dimensional Hilbert
space concomitant with operators interrelationships. We now turn to its possible use in the
mappings of Hilbert space operators onto the phase space like points and lines of the
geometry.

\section{Mapping onto phase space}

We now define a mapping of Hilbert space operators, e.g. an arbitrary operator, B, onto
the phase space - like lines of DAPG. The mapping is defined by  \cite{revzen2},
\begin{equation}
B\Rightarrow V(j;B)\equiv tr BP_j.
\end{equation}
Here $P_j$ is a line operator within DAPG. (Alternatively we could have cast the mappings within
APG as is clear from the discussion in the previous section.)
The density operator may be expressed in terms of $V(j;\rho)$:
\begin{equation}
\rho=\frac{1}{d}\sum^{d^2}_j\big(tr\rho P_j\big)P_j.
\end{equation}
It can be shown that  $V(j;\rho)$ plays the role of quasi distribution \cite{revzen2} in the phase space like lines of DAPG. Thus for example the expectation value of an arbitrary operator $B$ we have,
\begin{equation}
tr\rho B=\frac{1}{d}\sum^{d^2}V(j;\rho)V(j;B).
\end{equation}
The quasi distribution may be reconstructed from the expectation values of the point operator
 $A_{\alpha}$ i.e. MUB state projector's expectation value (obtained, e.g. by measurements),
\begin{equation}
tr \rho A_{\alpha}=\frac{1}{d}\sum^{d^2}V(j;\rho)V(j;A_{\alpha})=
\frac{1}{d}\sum^{d^2}V(j;\rho)\Lambda_{\alpha,j}.
\end{equation}
Thence
\begin{equation}
\frac{1}{d}\sum_{\alpha\in\alpha'\cup M_{\alpha'}}\sum^{d^2}V(j;\rho)\Lambda_{\alpha,j}=V(j;\rho).
\end{equation}

\section{Summary and Concluding Remarks}

It is of interest that, if we associate the CB states with the position variable, q, of the
continuous problem and its Fourier transform state, viz b=0
 (cf. Eq.(\ref{mxel})), with the momentum, p, we have that the line of the finite dimension problem
 is parameterized with "initial" values of "q" and "p" i.e. m(-1) and m(0).\\

Finite geometry stipulates interrelations among lines and points. The stipulations for the
(finite) dual affine plane geometry (DAPG) was shown to conveniently accommodate
association of geometric lines and points with projectors of states of mutual unbiased
bases (MUB). The latter act in a (finite dimensional, d) Hilbert space. This underpinning
of Hilbert space operator with DAPG reveal some novel inter operators relations. Noteworthy
among these are Hilbert space operators, $P_j,\;j=1,2,...d^2$, which are underpinned with
DAPG lines, $L_j$ that abide by $P_j^2=\hat{I}\;\forall j$, and are mutually orthogonal,
$tr P_jP_{j'}=d\delta_{j,j'}.$ These allow their utilization for general mapping of Hilbert
space operators onto the phase space like lines and points of DAPG in close analogy with
the mappings within the continuum of Hilbert space operators onto phase space via the well
known Wigner function \cite{moyal,ellinas}. If we associate the computational basis (CB)
states with the position variable, q, of the continuous problem, and its Fourier transform
state  (i.e. states of the basis that is diagonal for  translation operator, X
 (cf. Eq.(\ref{mxel})) with the momentum, p, we have that the line of the finite dimension problem
 is parameterized with these phase space like variables. We present a transcription from DAPG  to
 (finite) affine plane geometry (APG) underpinnings. Within the latter such labelling is natural
 and, further more, the line operators here include points that are aligned in a straight line.
 This interpret the APG operator underpinning that is given in \cite{wootters3} as due to
 the association of MUB state projectors with points within DAPG.

\section*[Appendix A]{Appendix: Fluctuation Distillation Formula}

Given,  Eq(34),   $\hat{P_j}=\sum_{\alpha \in j}\hat{A_\alpha}-\hat{I}$ and,
  Eq(\ref{p2}), $\hat{P_j}^2=\hat{I},$ implies

$$\big( \sum_{\alpha \in j}\hat{A_\alpha}-\hat{I}\big) \big(\sum_{\alpha' \in j}\hat{A_\alpha'}
-\hat{I} \big)=\hat{I}.$$ Thus,
$$\sum_{\alpha,\alpha' \in j}\hat{A_\alpha}\hat{A_\alpha'}=2\sum_{\alpha \in j}\hat{A_\alpha}.$$
Recalling that, Eq(\ref{ptop}), $A_{\alpha}^2=A_{\alpha}$ allows
$$\sum_{\alpha \ne\alpha' \in
j}\hat{A_\alpha}\hat{A_\alpha'}=\sum_{\alpha \in j}\hat{A_\alpha}.$$ QED

\end{document}